\documentclass{article}

\title{Formalization of dependent type theory: The example of
  \CaTT{}} 

\usepackage{macros} \usepackage{mathpartir} \usepackage{tikz-cd}
\usepackage{authblk}
\usepackage{hyperref}
\usepackage{amsmath,amsthm,amssymb,amsfonts}
\usepackage{geometry}

\newtheorem{theorem}{Theorem}
\newtheorem{proposition}[theorem]{Proposition}

\theoremstyle{definition}
\newtheorem{definition}[theorem]{Definition}

\theoremstyle{remark}
\newtheorem{example}[theorem]{Example}

\author[1]{Thibaut Benjamin} \affil[1]{Université Paris-Saclay, CEA,
  List, F-91120, Palaiseau, France, thibaut.benjamin@cea.fr}

\usepackage{newunicodechar}
\newunicodechar{∀}{\ensuremath{\mathnormal\forall}}
\newunicodechar{Γ}{\ensuremath{\mathnormal\Gamma}}
\newunicodechar{Δ}{\ensuremath{\mathnormal\Delta}}
\newunicodechar{Θ}{\ensuremath{\mathnormal\Theta}}
\newunicodechar{γ}{\ensuremath{\mathnormal\gamma}}
\newunicodechar{δ}{\ensuremath{\mathnormal\delta}}
\newunicodechar{→}{\ensuremath{\mathnormal\to}}
\newunicodechar{∘}{\ensuremath{\mathnormal{\circ}}}
\newunicodechar{ℕ}{\ensuremath{\mathbb{N}}}
\newunicodechar{≡}{\ensuremath{\mathnormal{\equiv}}}
\newunicodechar{⊢}{\ensuremath{\mathnormal{\vdash}}}
\newunicodechar{∈}{\ensuremath{\mathnormal{\in}}}
\newunicodechar{∗}{\ensuremath{\mathnormal{\ast}}}
\newunicodechar{⇒}{\ensuremath{\mathnormal{\Rightarrow}}}
\newunicodechar{¬}{\ensuremath{\mathnormal{\neg}}}
\newunicodechar{Ξ}{\ensuremath{\mathnormal{\Xi}}}
\newunicodechar{θ}{\ensuremath{\mathnormal{\theta}}}
\newunicodechar{Σ}{\ensuremath{\mathnormal{\Sigma}}}
\newunicodechar{λ}{\ensuremath{\mathnormal{\lambda}}}
\newunicodechar{𝕊}{\ensuremath{\mathbb{S}}}
\newunicodechar{𝔻}{\ensuremath{\mathbb{D}}}
\newunicodechar{ℓ}{\ensuremath{\mathnormal{\ell}}}
\newunicodechar{↦}{\ensuremath{\mathnormal{\mapsto}}}
\newunicodechar{⊘}{\ensuremath{\mathnormal{\emptyset}}}
\newunicodechar{∙}{\ensuremath{\mathnormal{\cdot}}}
\newunicodechar{≤}{\ensuremath{\mathnormal{\leq}}}
\newunicodechar{₀}{\ensuremath{\mathnormal{_{0}}}}
\newunicodechar{◃}{\ensuremath{\mathnormal{\triangleleft}}}
\newunicodechar{∂}{\ensuremath{\mathnormal{\partial}}}
\newunicodechar{⁺}{\ensuremath{\mathnormal{^+}}}
\newunicodechar{⁻}{\ensuremath{\mathnormal{^{-}}}}
\newunicodechar{⊂}{\ensuremath{\mathnormal{\subset}}}
\newunicodechar{≗}{\ensuremath{\mathnormal{\overset{o}=}}}
\newunicodechar{ᵢ}{\ensuremath{\mathnormal{_{i}}}}
\newunicodechar{₁}{\ensuremath{\mathnormal{_{1}}}}
\newunicodechar{₂}{\ensuremath{\mathnormal{_{2}}}}
\newunicodechar{∪}{\ensuremath{\mathnormal{\cup}}}
\newunicodechar{ᵤ}{\ensuremath{\mathnormal{_{u}}}}

\begin{document}

\maketitle

\begin{abstract}
  We present the type theory \CaTT{}, originally introduced by Finster
  and Mimram to describe globular weak $\omega$-categories, and we
  formalise this theory in the language of homotopy type theory. Most
  of the studies about this type theory assume that it is well-formed
  and satisfy the usual syntactic properties that dependent type
  theories enjoy, without being completely clear and thorough about
  what these properties are exactly. We use the formalisation that we
  provide to list and formally prove all of these meta-properties,
  thus filling a gap in the foundational aspect. We discuss the key
  aspects of the formalisation inherent to the theory \CaTT{}, in
  particular that the absence of definitional equality greatly
  simplify the study, but also that specific side conditions are
  challenging to properly model. We present the formalisation in a way
  that not only handles the type theory \CaTT{} but also all the
  related type theories that share the same structure, and in
  particular we show that this formalisation provides a proper ground
  to the study of the theory \MCaTT{} which describes the globular
  monoidal weak $\omega$-categories. The article is accompanied by a
  development in the proof assistant \Agda{} to actually check the
  formalisation that we present.
\end{abstract}

\section{Introduction}
This article aims at presenting and formalising the foundations of a
class of dependent type theories; the theory \CaTT{} introduced by
Finster and Mimram~\cite{catt} being the motivating example.  This
dependent type theory is designed to encode a flavor of higher
categorical structures called weak $\omega$-categories, and its
semantics has been proved~\cite{benjamin2021globular} to be equivalent
to a definition of weak $\omega$-categories due to
Maltsiniotis~\cite{maltsiniotis} based on an approach introduced by
Grothendieck~\cite{pursuing-stacks}. However, none of the
aforementioned articles about the theory \CaTT{} provide a satisfying
investigation of the type theoretic foundations, and simply assume
that a lot of syntactic meta-properties are satisfied. This is not a
shortcoming of those articles, but common practice in the type theory
community since low-level descriptions are very lengthy and would make
the articles essentially impossible to read. Thus it is usually
accepted that such a description is possible and satisfies all the
usually needed meta-theoretic properties can be proven by induction.
Yet, there does not a general framework for presenting a dependent
type theory, that would enforce the existence of low-level foundations
together with the usual meta-theoretic properties. As a result, every
new dependent type theory has to either be very specific about its
foundations, or be vague enough and rely on the reader's ability to
infer the foundations and convince themselves that the meta-theoretic
properties can be proven. This prevents a lot of results to be built
from the ground up, and makes research about dependent type theory not
readily available to experts.

The question of the choice of theoretical grounds to present a
dependent type theory is also an interesting question in its own
rights. It has been a long-standing goal to define type theory within
type theory~\cite{chapman2009type}, and significant progress has been
met in this direction, but there are still open problems when it comes
to fully embracing proof-relevance. In particular, it has been
theorised that homotopy type theory (HoTT)~\cite{hottbook} should be
completely definable within itself, but providing such a definition is
still one of the main open problems of HoTT. Notably, this definition
raises an important \emph{coherence problem} that is reminiscent in
its nature of the problems encountered in defining higher
structures. Altenkirch and Kaposi~\cite{altenkirch2016type} have made
progress towards solving this issue by using quotient inductive
inductive types, a type theoretic construct whose semantics is not
completely understood.

In this article, we present a proof of concept for using the language
of HoTT as a meta-theory for defining and studying dependent type
theories. Our main objective is the theory \CaTT{}, so our
formalisation within HoTT also provide a much-needed foundation to
this theory. Moreover this theory is particularly simple, since it
does not have definitional equality, making it an ideal candidate to
focus on one challenge posed by the use of HoTT. We first give a quick
informal presentation of the theory \CaTT{}, in order to assert our
goal. This presentation relies on a simpler type theory called \Glob{}
which describes globular sets. We then discuss the formalisation of
this theory, along with its meta-theoretic properties. Next, we move
on to the formalisation of dependent type theories whose shapes are
described by the theory \Glob{}, that we call globular type theories,
and whose \CaTT{} is our primary example.  Finally we present the
theory \CaTT{} and show how it can be formalised in the framework of
globular type theories. We prove some of the interesting
meta-properties of this theory with a particular emphasis on the ones
that are important to understand its semantics and that are used in
other articles without full
proofs~\cite{catt,benjamin2021globular}. All our definitions and
proofs are accompanied by a formalisation in the theorem prover
\Agda{}\footnote{\url{https://github.com/thibautbenjamin/catt-formalization}},
(to be consistent with the language of HoTT, we also deactivated the
use of the axiom K in \Agda{}).

\section{Introduction to the theory \CaTT{}}

We present here the type theory \CaTT{} in order to motivate our
formalisation work. The introduction we provide here focuses mostly
on the syntactic aspects of the theory, but we also provide some
intuition to the semantics, to help situate the theory in a broader
picture. We refer the reader to existing articles~\cite{catt,
  benjamin2021globular} for more in-depth discussions about the
semantics of this type theory. The first presentation we provide here
is informal, and serves as a guideline for the foundations that we are
developing.

\subsection{General setup for dependent type theories}
All along this article, we only consider dependent type theories which
support the contraction, exchange and weakening rules, so we simply
refer to them as type theories and assume those rules
implicitly. Those theories are centred around four kinds of object,
that we introduce here along with corresponding notations
\[
  \begin{array}{r@{\ :\ }l@{\qquad\quad}r@{\ :\ }l}
    \text{contexts} & \Gamma, \Delta, \ldots & \text{types} & A, B, \ldots \\
    \text{substitutions} & \gamma, \delta, \ldots & \text{terms} & t, u , \ldots\\
    \multicolumn{4}{l}{\text{Each of these object is associated to a well-formedness judgement}} \\
    \text{$\Gamma$ is a valid context} & \Gamma\vdash & \text{$A$ is a valid type $\Gamma$} & \Gamma\vdash A \\
    \text{$\gamma$ is a valid substitution from $\Delta$ to $\Gamma$} & \Delta\vdash\Gamma & \text{$t$ is a term of type $A$ in $\Gamma$} & \Gamma\vdash t:A
  \end{array}
\]
The type theories we are interested in do not have definitional
equalities, so these are the only judgements we do consider here.

\subsection{The theory \Glob{}}
We first introduce the theory \Glob{} which is simpler than the theory
\CaTT{} and serves as a basis on which this theory relies. In the
theory \Glob{} there are no term constructors, hence the only terms
are the variables. There are two type constructors, that we denote
$\Obj$ and $\Hom{}{}{}$. Those two constructors are subject to the
following introduction rules
\begin{align*}
  \inferrule{\null}{\emptyset\vdash}{\regle{ec}}
  & &
      \inferrule{\Gamma\vdash A}{\Gamma,x:A\vdash}{\regle{ce}} \quad \text{Where $x\notin\Var\Gamma$}\\
  \inferrule{\Gamma\vdash}{\Gamma\vdash\Obj}{\regle{$\Obj$-intro}} & & \inferrule{\Gamma\vdash A \\ \Gamma\vdash t:A~\\ \Gamma\vdash u:A}{\Gamma\vdash\Hom Atu}{\regle{$\Hom{}{}{}$-intro}}\\
  \inferrule{\Gamma\vdash \\ (x : A) \in \Gamma}{\Gamma\vdash x:A}{\regle{var}} \\
  \inferrule{\Delta\vdash}{\Delta\vdash \langle\rangle:\emptyset}{\regle{es}} & &\inferrule{\Delta\vdash \gamma: \Gamma \\ \Gamma,x:A\vdash~\\ \Delta\vdash t:A[\gamma]}{\Delta\vdash\langle\gamma,x\mapsto t\rangle:(\Gamma,x:A)}{\regle{se}}\\
\end{align*}
Where $t[\gamma]$ denotes the application of the substitution $\gamma$
to the term $t$ and $\Var{}$ denotes the set of variables needed to
write a syntactic object. The contexts of the theory \Glob{} can be
represented by \emph{globular sets}. Those are analogues to graph,
except that they are allowed to have arrows in every dimension (often
called \emph{cells}). A cell of dimension $n+1$ has for source and
target a pair of cells of dimension $n$ which are required to share
the same source and target. In the type theory, this is imposed by
rule~\regle{$\Hom{}{}{}$-intro} in which the terms $t$ and $u$ have to
share the type $A$.

\begin{example}\label{ex:glob-ctx}
  We illustrate the correspondence between contexts and finite
  globular sets with a few examples, using a diagrammatic
  representation of globular sets where we give the same name to a
  variable in a context and its corresponding cell in the
  corresponding globular set.
  \[
    \begin{array}{r@{\qquad}c}
      \Gamma_{c} = (x:\Obj,y:\Obj,f:\Hom \Obj xy,z:\Obj, h:\Hom\Obj yz) &
                                                                          \begin{tikzcd}[ampersand replacement=\&]
\overset{x}{\bullet} \ar[r,"f"] \& \overset{y}{\bullet} \ar[r, "h"] \& \overset{z}{\bullet}
\end{tikzcd}
      \\
      \Gamma_{w} = (x:\Obj,y:\Obj,f:\Hom \Obj xy ,g:\Hom \Obj xy, \alpha:\Hom{\Hom \Obj xy}fg, z:\Obj, h:\Hom\Obj yz) &
                                                                                                                        \begin{tikzcd}[ampersand replacement=\&]
\overset{x}{\bullet} \ar[r,"f", bend left]\ar[r,"g"', bend right]\ar[r,"\Downarrow", phantom] \& \overset{y}{\bullet} \ar[r, "h"] \& \overset{z}{\bullet}
\end{tikzcd}
      \\
      \Gamma_{\circlearrowleft} = (x:\Obj,f:\Hom\Obj xx) &
                                                           \begin{tikzcd}[ampersand replacement=\&]
\overset{x}{\bullet} \ar[loop right,"f"]
\end{tikzcd}
    \end{array}
  \]
\end{example}
This correspondence is not an actual bijection: Several context may
correspond to the same globular set, if they only differ by reordering
of the variables. One can account for this by considering the category
of contexts with substitutions. This category is equivalent to the
opposite of the category of finite globular
sets~\cite[Th. 16]{benjamin2021globular}. A judgement
$\Gamma\vdash x:A$ in the theory \Glob{} corresponds to a well-defined
cell $x$ in the globular set corresponding to $\Gamma$, and the type
$A$ provides all the iterated sources and targets of $x$.

\subsection{Ps-contexts}
In the type theory \CaTT{} there is an additional judgement on
contexts, that recognises a special class of contexts that we call
\emph{ps-contexts}. We denote this judgement $\Gamma\vdashps$, and we
introduce with the help of an auxiliary judgement
$\Gamma\vdashps x:A$. These two judgements are subject to the
following derivation rules
\begin{align*}
  \inferrule{\null}{(x:\Obj)\vdashps x:\Obj}{\regle{pss}}
  & & \inferrule{\Gamma\vdashps x:A}{\Gamma,y:A,f:\Hom Axy\vdashps f:\Hom Axy}{\regle{pse}} \\
  \inferrule{\Gamma\vdashps f:\Hom Axy}{\Gamma\vdashps y:A}{\regle{psd}}
  & &  \inferrule{\Gamma\vdashps x:\Obj}{\Gamma\vdashps}{\regle{ps}}
\end{align*}
The semantical intuition is that $\vdashps$ characterises the contexts
corresponding \emph{pasting schemes}~\cite{batanin1998monoidal,
  maltsiniotis}. Those are the finite globular sets defining an
essentially unique composition in weak $\omega$-categories. Finster
and Mimram given an alternate characterisation of the
ps-contexts~\cite{catt} using a relation on the variables of a
context, denoted $x\triangleleft y$. Each context $\Gamma$ defines
this relation as the transitive closure of the relation generated by
$x\triangleleft y\triangleleft z \text{ as soon as } \Gamma\vdash y
:\Hom{}xz \text{ is derivable.}$ The authors have proved that a
context is isomorphic to a ps-contexts if and only if this relation is
a linear order.

Additionally, a ps-context $\Gamma$ defines two subsets of its
variables $\src \Gamma$ and $\tgt \Gamma$ respectively called the
source and the target set.
\begin{definition}
  We define the \emph{$i$-source} of ps-context as a list by induction
  \[
    \src i {(x:\Obj)} = (x:\Obj)
    \qquad\partial^-_i(\Gamma,y:A,f:\Hom{}xy) = \left\{
      \begin{array}{l@{\quad}l}
        \partial^-_i\Gamma & \text{if $\dim A \geq i$}\\
        \partial^-_i\Gamma,y:A,f:\Hom{}xy & \text{otherwise}
      \end{array}
    \right.
  \]
  and its \emph{$i$-target} by
  \[
    \tgt i {(x:\Obj)} = (x:\Obj) \qquad
    \partial^+_i(\Gamma,y:A,f:\Hom{}xy) = \left\{
      \begin{array}{l@{\quad}l}
        \partial^+_i\Gamma & \text{if $\dim A > i$}\\
        \operatorname{drop}(\partial^+_i\Gamma),y:A & \text{if $\dim A = i$}\\
        \partial^+_i\Gamma,y:A,f:\Hom{}xy & \text{otherwise}
      \end{array}
    \right.
  \]
  where $\operatorname{drop}$ is the list with its head removed. The
  \emph{source} (resp. target) of a ps-context $\Gamma$ is the set of
  all variables in its $\dim\Gamma$-source (resp. in its
  $\dim\Gamma$-target).
\end{definition}
Semantically, these sets contain all the variables in the source and
target of the result in the application of the essentially unique
composition defined by the pasting scheme corresponding to $\Gamma$.

\begin{example}\label{ex:psc}
  The contexts $\Gamma_{c}$ and $\Gamma_{w}$ defined in
  Example~\ref{ex:glob-ctx} are ps-contexts, while the context
  $\Gamma_{\circlearrowleft}$ is not. We provide below a description
  of the relation $\triangleleft$ on those examples, as well as the
  source and target set variables for those that are ps-contexts.
  \[
    \begin{array}{c@{\quad :\quad }c@{\qquad}c@{\qquad}c}
      \Gamma_{c}
      & x\triangleleft f\triangleleft y\triangleleft h \triangleleft z
      & \src{\Gamma_{c}} = \set{x}
      & \tgt{\Gamma_{c}} = \set{z} \\
      \Gamma_{w}
      & x \triangleleft f \triangleleft \alpha \triangleleft g \triangleleft y \triangleleft h \triangleleft z
      & \src{\Gamma_{w}} = \set{x,f,y,h,z}
      & \tgt{\Gamma_{w}} = \set{x,g,y,h,z} \\
      \Gamma_{\circlearrowleft}
      &
        \begin{tikzcd}[ampersand replacement = \&,column sep = tiny]
x \ar[r, phantom, "\triangleleft", shift left]
\ar[r, phantom, "\triangleright", shift right]
\& f
\end{tikzcd}
    \end{array}
  \]
\end{example}

\subsection{The type theory \CaTT{}}
The type theory \CaTT{} is obtained from the type theory \Glob{} by
adding new term constructors that witness the operations of weak
omega-categories. There are two of these constructors, $\cohop$ and
$\coh$, in such a way that a term is either a variable or of the form
$\cohop_{\Gamma,A}[\gamma]$ or $\coh_{\Gamma,A}[\gamma]$, where in
both two last cases, $\Gamma$ is a ps-context, $A$ is a type and
$\gamma$ is a substitution. These term constructors are subject to the
following introduction rules.
\begin{align*}
  \inferrule{
  \Gamma\vdashps \\
  \Gamma\vdash t:A \\
  \Gamma\vdash u:A \\
  \Delta\vdash \gamma:\Gamma}
  {\Delta \vdash \cohop_{\Gamma,\Hom Atu}[\gamma] : \Hom{A[\gamma]}{t[\gamma]}{u[\gamma]}}{\regle{op}}
  & &
      \inferrule{
      \Gamma\vdashps \\
  \Gamma\vdash A \\
  \Delta\vdash \gamma:\Gamma}
  {\Delta \vdash \cohop_{\Gamma,A}[\gamma] : A[\gamma]}{\regle{coh}}
\end{align*}
Both these rules apply only under extra side-conditions. We denote
$\Vop$ the side condition of~\regle{op} and $\Vcoh$ the one
of~\regle{coh}. Those side conditions are the following
\[
  \Vop : \left\{
    \begin{array}{r@{=}l}
      \Var t \cup \Var A & \src\Gamma\\
      \Var u \cup \Var A & \tgt\Gamma
    \end{array}
  \right.  \qquad\qquad \Vcoh : \Var A = \Var \Gamma
\]
Recall that a ps-context is meant to represent an essentially unique
composition in a weak $\omega$-category. The rules~\regle{op}
and~\regle{coh} enforce this condition, in a weak sense analogue to
requiring the type of composition to be contractible in homotopy type
theory. More specifically, the rule~\regle{op} asserts that in a
context $\Delta$, for every situation described by a ps-context
$\Gamma$ (as witnessed by $\gamma$), there exists a term witnessing
the existence of the composition of this situation. The
rule~\regle{coh} imposes that any two such compositions are related by
a higher cells. Indeed in this rule the type $A$ is necessarily of the
form $\Hom{}{a}{b}$, where $a$ and $b$ represent two ways of composing
the ps-context. The role of the side condition is to prevent the
composition to apply partially.

\begin{example}
  Consider the context $\Gamma_{c}$ introduced in previous
  Example~\ref{ex:glob-ctx}, we have established in
  Example~\ref{ex:psc} that it is a ps-context, with
  $\src{\Gamma_{c}} = \set{x}$ and $\tgt{\Gamma_{c}} = \set{z}$. The
  type $\Hom\Obj xz$ thus satisfies the condition $\Vop$. This shows
  that for every context $\Delta$ with a substitution
  $\Delta\vdash\gamma:\Gamma_{c}$, we can define the term \texttt{comp
    $\gamma$}, and we have a derivation
  $\Delta\vdash\texttt{comp $\gamma$}:\Hom{}
  {x[\gamma]}{z[\gamma]}$. The substitution
  $\Delta\vdash\gamma:\Gamma_{c}$ picks out to composable arrows in
  $\Delta$, and the semantics is that the term \texttt{comp $\gamma$}
  witnesses the composition of those arrows. Additional examples are
  presented in~\cite{catt}.
\end{example}

Since there is no definitional equality or computation rule in this
theory, all the computational content is contained in the substitution
calculus, i.e., the action of substitutions on terms and types, and
the composition of substitutions.

\section{Structural foundation of dependent type theory}
In this section, we discuss the ways of presenting a dependent type
theory and motivate our preferred choice in the light of our objective
to formalise the theory \CaTT and its properties. There are
essentially two ways of approaching this formalisation problem, each
with its own strengths and weaknesses.

\subsection{The typed syntax approach}
An approach proposed by Dybjer~\cite{dybjer} for internalising the
semantics of dependent type theory is to define a typed syntax by
induction induction. In this approach, we define the syntax and the
judgements together, in such a way that every syntactic object
intrinsically comes with its well-formedness judgement. For this
reason we also refer to this approach as the \emph{intrinsic type
  theory}. In this setting it makes no sense to even consider an
ill-formed syntactic object, thus providing a very concise and
consistent presentation of the theory. This approach captures very
well the algebraic structure of dependent type theory. It also allows
for a variable-free presentation, avoiding dealing with technical
issues in managing the variable names (like accounting for
$\alpha$-equivalence). For all these reasons, an intrinsic approach to
dependent type theory is very valuable, the presentation is very
direct and reflects the structure of the theory we are formalising.

\subparagraph*{Structure of the typed syntax.}  The intrinsic
formalisation of dependent type theory relies on four types defined in
an inductive inductive way. They correspond to each of the four kinds
of syntactic objects in the theory. We do not define those fully here,
but still give their signature in \Agda{} pseudo-code in order to
discuss specific interesting points in following this route. We use
the keyword \verb|data| to introduce inductive or mutually inductive
types.
\[
  \begin{minipage}{.5\linewidth}
\begin{verbatim}
data Ctx : Set
data Sub : Ctx → Ctx
\end{verbatim}
  \end{minipage}
  \begin{minipage}{.5\linewidth}
\begin{verbatim}
data Ty : Ctx → Set
data Tm : ∀ Γ → Ty Γ → Set
\end{verbatim}
  \end{minipage}
\]
In this framework the type \verb|Ctx| is the type of well-formed
contexts \ie the type of all derivation trees of judgements of the
form $\Gamma\vdash$, and similarly for the three other types. This
excludes ill-formed objects from the syntax. For this reason, the type
of types \verb|Ty| depends on the type \verb|Ctx|: It does not make
sense to require a type to be well-formed, without a reference to the
context in which this well-formedness should be satisfied.

\subparagraph*{The added value of proof-assistants.}  Although it is
not apparent just by looking at the signature, the constructors of the
type \verb|Ctx| actually depend on the type \verb|Ty|, which itself
depends on the type \verb|Ctx| in its signature. Such hidden circular
dependencies are everywhere in the definition of those four
types. This means that checking the well-definedness of those types
and of every construction that we define by induction on those types
is non-trivial. As a result, using such an approach for a
pen-and-paper formalisation is very challenging. Not only do the
proofs have to be carried formally, but the proof that all the proofs
are well-formed should also be written explicitly. However, proof
assistants really can be a strong added-value in this situation, most
of them have a termination checking algorithm that handles the
well-formedness of each proof automatically for the user. The only
downside is that since the halting problem is undecidable, this
algorithm has to rely on heuristics, and forces the user to sometimes
over-specify some of the functions for the termination checking
algorithm to succeed.

\subparagraph*{The dependency problem.}  An intrinsic approach to
formalising dependent type theory presents hard challenges to solve
For instance, one of the first equality that we want to manipulate is
that the action of substitutions on types and terms respects the
composition. This translates to the following equalities
\[
  A[\gamma][\delta] = A[\gamma\circ\delta] \qquad\qquad
  t[\gamma][\delta] = t[\gamma\circ\delta]
\]
Within the typed syntax, the second of these equations is not even a
valid statement as is. Indeed, suppose given the term
\verb|t : Tm Γ A| and two substitutions \verb|γ : Sub Δ Γ| and
\verb|δ : Sub Θ Δ| then the term \verb|t[γ ∘ δ]| is of type
\verb|Tm Θ A[γ ∘ δ]|, whereas the term \verb|t[γ][δ]| is of type
\verb|Tm Θ A[γ][δ]|. Although these two types are equal by application
of the two equalities, they are not definitionally so. Thus to even
state this equality, one has to transport one of the terms along the
equality between the types. Proving later properties then requires a
proper handling of these transports to show that they agree, and these
new terms themselves have to be handled, and so on leading to a
coherence problem which would require the introduction of infinitely
many axioms to describe the theory as we want it. A solution to this
issue was introduced by Chapman~\cite{chapman2009type}, by allowing
the equality to compare terms that live in different types. This idea
was adapted by
Lafont~\footnote{\url{https://github.com/amblafont/omegatt-agda/tree/2tt-fibrant}}
as a heterogeneous predicate used to avoid transports. However this
requires assuming the uniqueness of identity proofs (UIP), at least
for all types of the form \verb|Ty Γ|. This trivialises the coherence
problem, but breaks the computational behaviour of the theory: nothing
reduce past the UIP axiom. Even with this simplification, the
aforementioned development is beyond the capabilities of \Agda{}, and
requires deactivating its termination checker, which is one of the
motivation to use a proof-assistant in the first place.

\subparagraph*{Typed syntax using quotient inductive inductive types.}
Our overview of the intrinsic approach to dependent type theory would
not be complete without a discussion of the approach developed by
Altenkirch and Kaposi~\cite{altenkirch2016type} with quotient
inductive inductive types. The intuition is to allow assumptions that
are higher order constructors of higher inductive inductive types, and
they solve the coherence problem by leveraging the power of higher
induction. They also truncate all the higher constructors to level $1$
(hence the name ``quotient''). However, quotient inductive inductive
types are not well understood semantically, and most modern proof
assistants do not support them naively. As a result, formalising this
approach in a proof assistant requires defining manually custom
recursors for the higher inductive types. Specifically in \Agda{},
these recursors directly conflict with the pattern-matching algorithm,
and thus a formalisation in this setting has to carefully avoid the
pattern-matching mechanism when it is inconsistent, and rely on user
defined methods. Those do not scale up very well, and the lack of
native facilities quickly become a practical obstruction to any
realistic non-trivial results about the theory.

\subsection{The raw syntax approach}
In this article, we take an alternate route to formalising dependent
type theory, based on the separation of the syntax and the rules of
the theory. To highlight the difference we call \emph{raw syntax} the
syntactic elements that are not tied to a derivation tree. Contrary to
the typed syntax, raw syntax may contain ill-formed entities that do
no correspond to any entity of the theory. In this approach, we
delegate the computational duty to the raw syntax, which completely
sidesteps the coherence problem. However this approach does not give a
direct description of the theory, it requires combining two separate
ingredients, the raw syntax and the judgements, to describe the
theory. For this reason, we call this approach an \emph{extrinsic}
formalisation of a dependent type theory. In this article we present a
formalisation of the dependent type theory \CaTT{} in an extrinsic
way. Similar formalisation's have been developed, notably by
Finster\footnote{\url{https://github.com/ericfinster/catt.io/tree/master/agda}}
and Rice\footnote{\url{https://github.com/alexarice/catt-agda}}, but
with the intent of proving other kinds of theories. Moreover Lafont,
Hirshowitz and Tabareau have also developed a formalisation of a type
theory related to \CaTT{} in a similar fashion in
\Coq{}\footnote{\url{https://github.com/amblafont/weak-cat-type/tree/untyped2tt}}~\cite{lafont2018types}
with the goal of internalising the observation that types are weak
$\omega$-groupoids~\cite{lumsdaine2009weak, van2011types,
  altenkirch2012syntactical}.

From now on we present our formalisation, and we use a lot of \Agda{}
pseudo-code. To simplify the notations, we use the convention that all
the free variables are implicitly universally quantified. Moreover, we
provide explanation for the unconventional syntax that we use, so that
readers not familiar with \Agda{} can read the article and the
supplementary formalisation the same way.

\subparagraph*{Variables management.}  In order to compute operations
on the raw syntax, we keep track of the variables to encode the
information present in the typing but not in the raw syntax. From a
semantics point of view, the names of the variables should be
irrelevant and we prefer to always work up to $\alpha$-equivalence. To
avoid dealing with these issues, we develop a foundation in which the
variables are natively normalised. Although each variable has a name,
this name is uniquely determined by the rules of the theory. Thus,
there is no possibility of renaming the variables and no quotient is
needed. We achieve this with a variation on De Bruijn levels: We
consider the type of variables to be the type of natural numbers
$\mathbb{N}$ and we require the contexts to enumerate their variables
in increasing order. Since there is no variable binder in the theory
\CaTT{}, this specification alone suffices to completely determine all
the variables in the theory.

\subparagraph*{Structure of the raw syntax.}  We now present the
foundational structure of the dependent type theories that we are
interested in. They are particular case of dependent type theories
that support the contraction, weakening and exchange. In order to
highlight the structure of the type theory itself and not get lost in
technical difficulties tied to a specific dependent type theory like
\Glob{} or \CaTT{}, we first present the rule for the empty dependent
type theory. It is the dependent type theory with no type constructors
or term constructors. Semantically, this is a vacuous theory, but it
still implements all the structure of dependent type theory. We first
define the raw syntax: Contexts are supported by lists of pairs of the
form $(x,A)$ where $x$ is a variable and $A$ is a type, substitutions
are supported by lists of pairs $(x,t)$ where $x$ is a variable and
$t$ is a term, since there are no term constructors, the only terms
are variables and since there is no type constructors, there is no
type.

\begin{definition}
  We define the raw syntax of the empty dependent type theory as a
  collection of four types defined by mutual induction, representing
  respectively the (raw) contexts, substitutions, terms and types
  \[
    \begin{minipage}{.5\linewidth}
\begin{verbatim}
  Pre-Ctx = list (ℕ × Pre-Ty)
  Pre-Sub = list (ℕ × Pre-Tm)
\end{verbatim}
    \end{minipage}
    \begin{minipage}{.5\linewidth}
\begin{verbatim}
  data Pre-Tm where
    Var : ℕ → Pre-Tm
  data Pre-Ty where
\end{verbatim}
    \end{minipage}
  \]
\end{definition}
Here, the constructor \verb|Var| produces an inhabitant of the type
\verb|Pre-Tm| from a variable (of type \verb|ℕ|), and there is no
constructor for the type \verb|Pre-Ty| since there is no type. Those
types do not need to be mutually inductive here, because we have not
added the constructors. When we add constructors later, these types
will become mutually inductive, and so we define them as such here for
consistency.

\subparagraph*{The action of substitutions.}  The computational
content of the theory is completely determined by the action of
substitutions and the composition of substitution. We define these
operations on the raw syntax levels, and they are the reason why we
need to introduce variable names. The coherence problem does not
appear at this level since none of the types constituting the raw
syntax are dependent. The composition of substitutions can be seen as
an action of substitutions on substitutions.
\begin{definition}
  We define the action of substitutions on types, terms and the
  composition of substitutions as the following mutually inductive
  operations
\begin{verbatim}
  _[_]T : Pre-Ty → Pre-Sub → Pre-Ty
  _[_]t : Pre-Tm → Pre-Sub → Pre-Tm
  _∘_ : Pre-Sub → Pre-Sub → Pre-Sub

  () [ γ ]T

  Var x [ nil ]Pre-Tm = Var x
  Var x [ γ :: (v , t) ]Pre-Tm = if x ≡ v then t else ((Var x) [ γ ]Pre-Tm)

  nil ∘ γ = nil
  (γ :: (x , t)) ∘ δ = (γ ∘ δ) :: (x , (t [ δ ]Pre-Tm))
\end{verbatim}
\end{definition}
Here the case \verb|()[ γ ]T| represents the empty case, since there
are no constructors for the type \verb|Pre-Ty|. Moreover, \verb|nil|
denotes the empty list, and \verb|_::_| is the cons operation on
lists. Again there is no need strictly speaking to use mutually
inductive types here, but we do for the sake of consistency with other
dependent type theories.

\subparagraph*{Judgements and inference rules.}  In order to recover
the actual dependent type theory from the raw syntax, we need to
eliminate the ill-formed syntactic entities and for this we introduce
the judgements together with their inference rules.  We define those
are mutually inductive types that depend on the raw syntax of the
theory with the following signature
\begin{verbatim}
 data _⊢C : Pre-Ctx → Set
 data _⊢T_ : Pre-Ctx → Pre-Ty → Set
 data _⊢t_#_ : Pre-Ctx → Pre-Tm → Pre-Ty → Set
 data _⊢S_>_ : Pre-Ctx → Pre-Sub → Pre-Ctx → Set
\end{verbatim}
The type constructors of these mutually inductive types are exactly
the inference rules of the theory. The type \verb|Γ ⊢C|, for instance,
is meant to represent the judgement $\Gamma\vdash$, an inhabitant of
this type is built out of the type constructors, corresponding to
inference rules. Hence an element of this type can be thought of as a
derivation tree. Reasoning by induction on derivation trees, a common
technique to prove meta-theoretic properties of dependent type theory,
just translates to reasoning by induction on those four types.  This
discussion holds because there is computation rules (\ie rules
postulating definitional equalities) in our theories: In the presence
of such rules, one would need to consider higher inductive types, and
the computation rules would be higher order constructors.
\begin{definition}
  We define the following inference rules for constructing contexts,
  substitutions and variable terms, which are the inference rules of
  the empty type theory (again, the type \verb|_⊢T_| has no
  constructor, since it is empty).
\begin{verbatim}
data _⊢C where
  ec : nil ⊢C
  cc : Γ ⊢C → Γ ⊢T A → x == length Γ → (Γ :: (x , A)) ⊢C
data _⊢T_ where
data _⊢t_#_ where
  var : Γ ⊢C → (x , A) ∈ Γ → Γ ⊢t (Var x) # A
data _⊢S_>_ where
  es : Δ ⊢C → Δ ⊢S nil > nil
  sc : Δ ⊢S γ > Γ → (Γ :: (x , A)) ⊢C → (Δ ⊢t t # (A [ γ ]Pre-Ty))
                  → x == y → Δ ⊢S (γ :: (y , t)) > (Γ :: (x , A))
\end{verbatim}
\end{definition}
For our presentation, we rely on the correspondence between the term
constructors of those types and the rules to name the constructors
accordingly to the names we have given for the rules. In the rules
\verb|cc| and \verb|sc|, we consider free variables and require
equalities between them, instead of considering syntactically equal
terms so that we are able to eliminate on this equality only when
needed. This lets us avoid the use of axiom K. In the rule \verb|cc|,
the condition \verb|x == length Γ| is the one enforcing our condition
that context enumerate their variables in order.

\section{Formalisation and properties of the theory \Glob{}}
From now on, we consider the type theoretic foundations that we have
presented of an extrinsic formalisation of dependent type theory.  We
present our first formalisation of an actual dependent type theory
with the theory \Glob{}. To this end, we extend our definition of the
empty type theory by adding two constructors for the type
\verb|Pre-Ty|, corresponding to the two type constructors of the type
theory \Glob{}. We then show some interesting meta-properties that the
theory \Glob{} enjoys.

\subsection{Formal presentation of the theory \Glob{}}
Recall that the type theory \Glob{} has only two type constructors
$\Obj$ and $\Hom{}{}{}$. We thus introduce two constructors to the
type \verb|Pre-Ty|. Note that this addition changes the entire raw
syntax, since the other types \verb|Pre-Ctx|, \verb|Pre-Tm| and
\verb|Pre-Sub| are all mutually inductively defined with
\verb|Pre-Ty|. For simplicity purposes, we only present here the parts
whose definition changes, but these changes actually propagate to the
entire raw syntax.
\begin{definition}
  The raw syntax of the type theory \Glob{} is defined the same way as
  the raw syntax of the empty type theory, replacing the type
  \verb|Pre-Ty| with the type
\begin{verbatim}
data Pre-Ty where
  ∗ : Pre-Ty
  ⇒ : Pre-Ty → Pre-Tm → Pre-Tm → Pre-Ty
\end{verbatim}
  Additionally, we introduce the action of substitution to be defined
  the same way on raw terms and substitutions, and defined on raw
  types by
\begin{verbatim}
∗ [ γ ]Pre-Ty = ∗
⇒ A t u [ γ ]Pre-Ty = ⇒ (A [ γ ]Pre-Ty) (t [ γ ]Pre-Tm) (u [ γ ]Pre-Tm)
\end{verbatim}
\end{definition}
We also specify the judgements of the theory, again those are defined
in the same way for the empty type theory, except for the type
judgement \verb|_⊢T_|. Since these judgements are mutually inductive,
a change here again propagates to all the judgements of the theory.
\begin{definition}
  The type theory \Glob{} is obtained, from its raw syntax by adding
  the judgements defined in the same way as for the empty type theory,
  except for the following
\begin{verbatim}
data _⊢T_ where
  ob : ∀ {Γ} → Γ ⊢C → Γ ⊢T ∗
  ar : ∀ {Γ A t u} → Γ ⊢t t # A → Γ ⊢t u # A → Γ ⊢T ⇒ A t u
\end{verbatim}
\end{definition}
With this formalisation of the theory \Glob{}, we can prove a lot of
interesting meta-properties that are required to study its
semantics. All of these properties are proved by induction on the
derivation trees. A lot of these proofs are straightforward and we
just give a discussion on the non-trivial proof techniques that come
up. We refer the reader to the \Agda{} implementation provided as
supplementary material for the complete proofs of all of these
properties.

\subsection{Structure of the dependent type theory}
First we show a few properties about the theory \Glob{}, ensuring that
the raw syntax together with a typing rule describe a dependent type
theory with all the expected structure. Although these results are
used a lot to study the semantics of the
theory~\cite{catt,benjamin2021globular}, they are generally admitted
and proving them requires as much specificity on the foundational
aspects as we have provided here.

\subparagraph*{Weakenings and judgement preservation.}
We have claimed that the type theories we formalise implement the
weakenings, contractions and exchange rules. Our naming convention for
the contexts actually prevent two variables to have the same name, so
all our contexts are reduced with respect to contraction and the
contraction rule trivial. Exchange is a bit difficult to express due
to the dependency, and is not useful for future applications, so we do
not prove it here. The following proposition shows, among other things
the weakenings.
\begin{proposition}\label{prop:glob-syntactic}
  The theory \Glob{} supports weakenings for types, terms and
  substitutions.
\begin{verbatim}
wkT : Γ ⊢T A → (Γ :: (y , B)) ⊢C → (Γ :: (y , B)) ⊢T A
wkt : Γ ⊢t t # A → (Γ :: (y , B)) ⊢C → (Γ :: (y , B)) ⊢t t # A
wkS : Δ ⊢S γ > Γ → (Δ :: (y , B)) ⊢C → (Δ :: (y , B)) ⊢S γ > Γ
\end{verbatim}
  Moreover, any sub-term of a derivable term is itself derivable. More
  precisely, we have the following (\cf admitted results~\cite[Lemma
  6]{catt} and~\cite[Lemma 6]{benjamin2021globular})
  \[
    \begin{minipage}{.5\linewidth}
\begin{verbatim}
Γ⊢A→Γ⊢ : Γ ⊢T A → Γ ⊢C
Γ⊢t:A→Γ⊢ : Γ ⊢t t # A → Γ ⊢C
Δ⊢γ:Γ→Γ⊢ :  Δ ⊢S γ > Γ → Γ ⊢C
Δ⊢γ:Γ→Δ⊢ :  Δ ⊢S γ > Γ → Δ ⊢C
\end{verbatim}
    \end{minipage}
    \begin{minipage}{.5\linewidth}
\begin{verbatim}
Γ,x:A⊢→Γ⊢ : (Γ :: (x , A)) ⊢C → Γ ⊢C
Γ⊢t:A→Γ⊢A : Γ ⊢t t # A → Γ ⊢T A
Γ⊢src : Γ ⊢T ⇒ A t u → Γ ⊢t t # A
Γ⊢tgt : Γ ⊢T ⇒ A t u → Γ ⊢t u # A
\end{verbatim}
    \end{minipage}
  \]
\end{proposition}

\subparagraph*{Structure of category with families.}  Categories with
families~\cite{dybjer} (CwF) are a categorical framework designed to
describe the structure of a dependent type theory. They are ubiquitous
in the study of the semantics of dependent type theory.  With the
foundations that we have given, we can prove by induction on the
derivation trees that the type theory \Glob{} defines a CwF. However,
developing category theory within HoTT can prove quite challenging,
and to avoid doing so, we prove the separately all the required
ingredients that constitute a CwF.
\begin{proposition}\label{prop:action}
  The action of a derivable substitution on a derivable type (\resp{}
term) is again a derivable type (\resp{} term)
\begin{verbatim}
[]T : Γ ⊢T A → Δ ⊢S γ > Γ → Δ ⊢T (A [ γ ]Pre-Ty)
[]t : Γ ⊢t t # A → Δ ⊢S γ > Γ → Δ ⊢t (t [ γ ]Pre-Tm) # (A [ γ ]Pre-Ty)
\end{verbatim}
\end{proposition}

\begin{proposition}\label{prop:id}
  In the theory \Glob{}, there is an identity substitution defined by
\begin{verbatim}
Pre-id : ∀ (Γ : Pre-Ctx) → Pre-Sub
Pre-id nil = nil
Pre-id (Γ :: (x , A)) = (Pre-id Γ) :: (x , Var x)
\end{verbatim}
  It acts trivially on types, terms and substitutions on a raw syntax
  level and it is derivable.
  \[
    \begin{minipage}{.5\linewidth}
\begin{verbatim}
[id]T : (A [ Pre-id Γ ]Pre-Ty) == A
[id]t : (t [ Pre-id Γ ]Pre-Tm) == t
\end{verbatim}
    \end{minipage}
    \begin{minipage}{.5\linewidth}
      \begin{verbatim}
∘-right-unit : (γ ∘ Pre-id Δ) == γ
Γ⊢id:Γ : Γ ⊢C → Γ ⊢S Pre-id Γ > Γ
\end{verbatim}
    \end{minipage}
  \]
\end{proposition}

\begin{proposition}\label{prop:action-comp}
  The action of substitution is compatible with the composition of
  substitution
\begin{verbatim}
[∘]T : Γ ⊢T A → Δ ⊢S γ > Γ → Θ ⊢S δ > Δ →
   ((A [ γ ]Pre-Ty) [ δ ]Pre-Ty) == (A [ γ ∘ δ ]Pre-Ty)
[∘]t : Γ ⊢t t # A → Δ ⊢S γ > Γ → Θ ⊢S δ > Δ →
   ((t [ γ ]Pre-Tm) [ δ ]Pre-Tm) == (t [ γ ∘ δ ]Pre-Tm)
\end{verbatim}
\end{proposition}

\begin{proposition}\label{prop:comp}
  The composition of substitutions preserves the derivability of
  substitutions. Moreover, it is associative, and the identity is the
  left unit of the composition.
\begin{verbatim}
∘-adm : Δ ⊢S γ > Γ → Θ ⊢S δ > Δ → Θ ⊢S (γ ∘ δ) > Γ
a : Δ ⊢S γ > Γ → Θ ⊢S δ > Δ → Ξ ⊢S θ > Θ → (γ ∘ δ) ∘ θ == γ ∘ (δ ∘ θ)
∘-left-unit : Δ ⊢S γ > Γ → (Pre-id Γ ∘ γ) == γ
\end{verbatim}
\end{proposition}
The result in Proposition~\ref{prop:action-comp} and~\ref{prop:comp}
may not hold at the raw syntax level, and they do require derivability
hypothesis to hold. All these propositions together show a large part
of the structure of CwF of the theory \Glob{}. We have presented them
in an unusual order which highlights their dependency structure: Each
of the later cited result uses the previously cited ones. These
results together correspond to~\cite[Propositions 8, 9 and
10]{benjamin2021globular}, where they are used without proof.

\subsection{Proof-theoretic considerations}
In our foundational framework, we can also show meta-properties of a
proof-theoretic nature. From now on we will use the language of HoTT
as a convenience, even though we do not need the univalence axiom, and
bare Martin-Löf type theory without any axiom is sufficient.

\subparagraph*{Decidability of type checking.}
In Martin-Löf type theory, we express that a type \verb|A| is
decidable, by exhibiting an inhabitant of the type
\verb|dec A = A + ¬ A|.
\begin{theorem}\label{thm:dec-type-checking}
  Type checking in the theory \Glob{} is a decidable problem
  \[
    \begin{minipage}{.51\linewidth}
\begin{verbatim}
dec-⊢C : ∀ Γ → dec (Γ ⊢C)
dec-⊢S : ∀ Δ Γ γ → dec (Δ ⊢S γ > Γ)
\end{verbatim}
    \end{minipage}
    \begin{minipage}{.49\linewidth}
\begin{verbatim}
dec-⊢T : ∀ Γ A → dec (Γ ⊢T A)
dec-⊢t : ∀ Γ A t → dec (Γ ⊢t t # A)
\end{verbatim}
    \end{minipage}
  \]
\end{theorem}
The proof of this theorem is by mutual induction of the derivation
tree, however the structure of the induction is quite complicated and
showing that it is well-founded is a hard problem. This is where the
use of a proof assistant like \Agda{} becomes extremely useful, since
its termination checker is able to verify this automatically. Proving
this theorem in \Agda{} amounts to implementing a certified type
checker for the theory \Glob{}.

\subparagraph*{Uniqueness of derivation tree.}
In order to express uniqueness, we use the language of HoTT, and we
define the types
\begin{verbatim}
is-contr A = Σ A (λ x → ((y : A) → x == y))
is-prop A = ∀ (x y : A) → is-contr (x == y)
\end{verbatim}
witnessing respectively that a type \verb|A| has a unique inhabitant,
and that \verb|A| is either empty or has a unique inhabitant.
\begin{theorem}\label{thm:uniqueness-derivation}
  In the theory \Glob{}, every derivable judgement has a unique
  derivation (stated without proof~\cite[Lemma
  7]{benjamin2021globular})
  \[
    \begin{minipage}{.5\linewidth}
\begin{verbatim}
is-prop-⊢C : is-prop (Γ ⊢C)
is-prop-⊢S : is-prop (Δ ⊢S γ > Γ)
\end{verbatim}
    \end{minipage}
    \begin{minipage}{.5\linewidth}
\begin{verbatim}
is-prop-⊢T : is-prop (Γ ⊢T A)
is-prop-⊢t : is-prop (Γ ⊢t t # A)
\end{verbatim}
    \end{minipage}
  \]
\end{theorem}
We again prove by mutual induction, and the proof is fairly
straightforward. The key ingredient is that the theory does not have
definitional equality.  We can recover the typed syntax from the raw
syntax and the judgements, by considering dependent pairs of a an
element of the raw syntax together with its judgement as follows. For
instance for contexts, we define the type
\verb|Ctx = Σ Pre-Ctx (λ Γ → Γ ⊢C)|, and similarly for types, terms
and substitutions, we define the types \verb|Ty Γ|, \verb|Tm Γ A| and
\verb|Sub Δ Γ|.

\subsection{Familial representability of types}
We now use our foundational framework to define the disks and sphere
contexts, which are families of contexts that play an important in the
understanding of the semantics of the
theory~\cite{benjamin2021globular}.
\begin{definition}\label{def:glob-disk}
  For every number \verb|n|, we define a type \verb|⇒ᵤ n| (u stands
  for ``universal'') and two contexts \verb|Pre-𝕊 n| and
  \verb|Pre-𝔻 n| by mutual induction in the raw syntax as follows
  (where \verb|ℓ| is the length of the list)
\begin{verbatim}
⇒ᵤ O = ∗
⇒ᵤ (S n) = ⇒ (⇒ᵤ n) (Var (2 n)) (Var (2 n + 1))
𝕊 O = nil
𝕊 (S n) = (𝔻 n) :: (ℓ (𝔻 n) , ⇒ᵤ n)
𝔻 n = (𝕊 n) :: (ℓ (𝕊 n) , ⇒ᵤ n)
\end{verbatim}
\end{definition}

\begin{proposition}\label{prop:disk}
  The disk and sphere contexts are valid contexts in the theory
  \Glob{}, and the type \verb|⇒ᵤ| is derivable in the sphere context
  \[
    \begin{minipage}{.32\linewidth}
\begin{verbatim}
𝕊⊢ : ∀ n → 𝕊 n ⊢C
\end{verbatim}
    \end{minipage}
    \begin{minipage}{.32\linewidth}
\begin{verbatim}
𝔻⊢ : ∀ n → 𝔻 n ⊢C
\end{verbatim}
    \end{minipage}
    \begin{minipage}{.36\linewidth}
\begin{verbatim}
𝕊⊢⇒ : ∀ n → 𝕊 n ⊢T ⇒ᵤ n
\end{verbatim}
    \end{minipage}
  \]
\end{proposition}
The sphere contexts play a particular role in the theory since they
classify the types in a context: types in a context are equivalent to
substitution from that context to a disk context. This is a result
that we call familial representability of
types~\cite{benjamin2021globular}, and that we formally prove in our
foundational framework, using the definition of equivalence
\verb|is-equiv| usual to HoTT.
\begin{theorem}\label{thm:up-disk}
  For every context $\Gamma$, and any derivable substitution from
  $\Gamma$ to a sphere, we define a derivable type in $\Gamma$ by
  applying the substitution on the type \verb|⇒ᵤ _|. The resulting map
  defines an equivalence
\begin{verbatim}
Ty-n : ∀ Γ → Σ ℕ (λ n →  Sub Γ (𝕊 n)) → Ty Γ
Ty-n Γ (n , (γ , Γ⊢γ:Sn) ) = ((⇒ᵤ n)[ γ ]Pre-Ty) , ([]T (𝕊⊢⇒ n) Γ⊢γ:Sn)

Ty-classifier : ∀ Γ → is-equiv (Ty-n Γ)
\end{verbatim}
\end{theorem}
This result is substantially harder to prove formally than the
previously mentioned ones, and relies on the uniqueness of derivation
trees.

\section{Formalisation and properties of the theory \CaTT{}}

In this section we adapt our foundational framework to define and
study the properties of the theory \CaTT{}. This requires the
introduction of the two term constructors $\cohop$ and $\coh$.  We
break down this work in two steps: First, we define a notion of
\emph{globular type theory} - is a dependent type theory with the same
type constructors as \Glob{} and generically indexed term constructors
- and we extend the meta-theoretic results of the previous section all
these theories. Then we define the type theory \CaTT{} as a particular
instance of a globular type theory, by specifying the index for the
term constructors.
\subsection{Globular type theories}
Globular type theories are dependent type theories that have the same
type structure as the theory \Glob{}, but have term constructors. In
order to describe not only the type theory \CaTT{}, but also other
dependent type theories, we define these term constructors
generically. To this end, we assume a type \verb|I|, which serves as
an index to all the term constructors.
\begin{definition}
  The raw syntax of a globular type theory is defined by the four
  mutually inductive types.
\begin{verbatim}
data Pre-Ty where
  ∗ : Pre-Ty
  ⇒ : Pre-Ty → Pre-Tm → Pre-Tm → Pre-Ty
data Pre-Tm where
  Var : ℕ → Pre-Tm
  Tm-c : ∀ (i : index) → Pre-Sub → Pre-Tm
data Pre-Sub where
  <> : Pre-Sub
  <_,_↦_> : Pre-Sub → ℕ → Pre-Tm → Pre-Sub
data Pre-Ctx where
  ⊘ : Pre-Ctx
  _∙_#_ : Pre-Ctx → ℕ → Pre-Ty → Pre-Ctx
\end{verbatim}
\end{definition}
Due to the mutually inductive definition, raw contexts and
substitutions, cannot be defined as mere lists. However, they behave
the exact same way as lists, and for all intents and purposes, we
treat them as normal lists of pairs in the rest of this article.
\begin{definition}
  The action of substitution on the raw syntax is computed the same way
  as the action of substitutions on the raw syntax of the theory
  \Glob{}, except on terms where it is defined by
\begin{verbatim}
Var x [ <> ]Pre-Tm = Var x
Var x [ < δ , v ↦ t > ]Pre-Tm = if x ≡ v then t else ((Var x) [ δ ]Pre-Tm)
Tm-c i γ [ δ ]Pre-Tm = Tm-c i (γ ∘ δ)
\end{verbatim}
\end{definition}
Due to the added dependency, the composition of substitution is now
mutually defined with the action on types and terms, whereas in the
theory \Glob{}, it was separately defined.

\subparagraph*{Inference rules of globular type theories.}
We give a presentation of a generic form for the introduction of the
indexed term constructors in globular type theories. To achieve this,
we parameterise the rules, in such a way that every term constructor
corresponds to its own introduction rule. We allow to have term
constructors in the pre-syntax that do not correspond to any derivable
term, if the rule is inapplicable. From now on, we assume that the
type \verb|I| has decidable equality, that is, we have a term
\begin{verbatim}
eqdecI : ∀ (x y : I) → dec (x == y)
\end{verbatim}
In order to parameterise the rules, we suppose that for every
inhabitant \verb|i| of the type \verb|I|, there exists a context
\verb|Ci i| in the raw syntax of \Glob{} and a type \verb|Ti i| in the
raw syntax of the globular type theory. Moreover, we assume that the
context \verb|Ci i| is derivable in the theory \Glob{}.

\begin{definition}
  A globular type theory is a theory obtained from its syntax by
  imposing the same judgement rules as in the theory \Glob{} for
  contexts, types and substitutions, and imposing for terms
\begin{verbatim}
data _⊢t_#_ where
  var : Γ ⊢C → (x , A) ∈ Γ → Γ ⊢t (Var x) # A
  tm : Ci i ⊢T Ti i → Δ ⊢S γ > Ci i → Δ ⊢t Tm-c i γ # (Ti i [ γ ]Pre-Ty)
\end{verbatim}
\end{definition}
Note that again, the judgements of the theory are defined mutually
inductively, and this change propagates to the other types. In
practice, we have introduced a type \verb|A| freely along with the
hypothesis \verb|A == Ti i [ γ ]Pre-Ty| in order to eliminate on this
equality.

\subparagraph*{Properties of globular type theories.}
Most of the meta-theoretic properties extend from the theory \Glob{}
to any globular type theory, but there can be some difficulties in
doing so.
\begin{proposition}
  Every globular type theory satisfy all the results presented in
  Propositions~\ref{prop:glob-syntactic}, \ref{prop:action},
  \ref{prop:id}, \ref{prop:action-comp} and~\ref{prop:comp}
\end{proposition}
In this case, these results are a bit more involved to prove, because
of the added dependency of terms on substitutions. Many results that
could be proven separately in the case of \Glob{} now depend on each
other and have to be proven by mutual induction. Again, termination
checking is not trivial, this is one instance where using \Agda{} is a
strong benefit.
\begin{theorem}[\cf Theorem~\ref{thm:uniqueness-derivation}]
  In any globular type theory, every derivable judgement has a single
  derivation tree.
\end{theorem}
Definition~\ref{def:glob-disk} of the disks and sphere contexts also
makes sense in any globular type theory. We also call those the disk
and sphere contexts in the raw syntax of the globular theory.
\begin{theorem}[cf Proposition~\ref{prop:disk} and
  Theorem~\ref{thm:up-disk}]
  The disk and sphere context define valid contexts in any globular
  type theory, and the sphere contexts classify the types: There is an
  equivalence between the derivable types in a contexts and the
  substitutions from that context to a sphere context.
\end{theorem}

\subparagraph*{Decidability of type checking.}
The decidability of type checking is a result that does not generalise
as well to any globular type theory, because the generic form we have
given for the rules is too permissive. Trying to reproduce the proof
of \Glob{} yields a proof whose termination cannot be checked by
\Agda{}: There is not a variant that decreases along the rules. And
indeed, it is possible to devise a globular type theory for which type
checking is not decidable. However, we can restrict our attention a
little further, and consider theories that satisfies an extra
hypothesis
\begin{verbatim}
wfI : ∀ i → Ci i ⊢T Ti i → dimC (Ci i) ≤ dim (Ti i)
\end{verbatim}
where \verb|dim| is the dimension of a type (\ie the number of
iterated arrows it is built with) and \verb|dimC| is the dimension of
a context (\ie the maximal dimension among the types it contains). The
theory \CaTT{} satisfies this hypothesis.
\begin{theorem}[cf~Theorem~\ref{thm:dec-type-checking}]
  For every globular type theory satisfying the hypothesis \verb|wfI|,
  the type checking is decidable.
\end{theorem}
Proving this by induction is fairly straightforward in principle, but
ensuring that the induction is well-formed is quite involved. Indeed,
there is no obvious decreasing variant and the proof relies on keeping
track of both the dimension and the number of nested term constructors
in a precise way to exhibit one. This argument is really non-trivial
and for this result the use of a termination checker such as \Agda{}'s
is extremely valuable.

\subsection{Ps-contexts and the theory \CaTT{}}
We leverage the definition of globular type theory to formalise and
prove some meta-theoretic properties of the theory \CaTT{}. To this
end, we define a particular type \verb|J| to index the term
constructors, as well as the contexts \verb|Ci j| and the types
\verb|Ti j| to define the inference rules.

\subparagraph*{Ps-contexts.}
In our formalism, there is no specific difference between the term
constructors $\cohop$ and $\coh$, both of them are term constructors
of the form \verb|Tm-c|. If anything, formally, $\cohop$ and $\coh$
correspond to families of term constructors and not term
constructors. One of the ingredients to index these families are the
ps-contexts that we formally define here.
\begin{definition}
  We define the judgements \verb|_⊢ps| and \verb|_⊢_#_| over the raw
  syntax of the type theory \Glob{} as the following inductive types
  (where we denote \verb|l| for \verb|ℓ Γ|)
\begin{verbatim}
data _⊢ps_#_ : Pre-Ctx → ℕ → Pre-Ty → Set where
  pss : (nil :: (O , ∗)) ⊢ps O # ∗
  psd : Γ ⊢ps f # (⇒ A (Var x) (Var y)) → Γ ⊢ps y # A
  pse : Γ ⊢ps x # A  → ((Γ :: (l , A)) :: (S l , ⇒ A (Var x) (Var l))) ⊢ps
       S l # ⇒ A (Var x) (Var l)

data _⊢ps : Pre-Ctx → Set where
  ps : Γ ⊢ps x # ∗ → Γ ⊢ps
\end{verbatim}
\end{definition}
\begin{proposition}
  The ps-contexts are valid contexts of the theory \Glob.
\begin{verbatim}
Γ⊢ps→Γ⊢ : Γ ⊢ps → Γ ⊢
\end{verbatim}
\end{proposition}

\subparagraph*{The relation $\triangleleft$.}
To work with ps-contexts formally, we define the relation
$\triangleleft$ introduced by Finster and Mimram~\cite{catt}. The main
purpose of this relation is to perform inductive reasoning.
\begin{definition}
  Given a contest $\Gamma$, we define a generating relation
  \verb|Γ ,_◃₀_|, together with its transitive closure \verb|Γ ,_◃_|
  as follows
\begin{verbatim}
data _,_◃₀_ Γ x y : Set where
  ◃∂⁻ : Γ ⊢t (Var y) # (⇒ A (Var x) (Var z)) → Γ , x ◃₀ y
  ◃∂⁺ : Γ ⊢t (Var x) # (⇒ A (Var z) (Var y)) → Γ , x ◃₀ y

data _,_◃_ Γ x y : Set where
  gen : Γ , x ◃₀ y → Γ , x ◃ y
  ◃T :  Γ , x ◃ z → Γ , z ◃₀ y → Γ , x ◃ y
\end{verbatim}
\end{definition}
\begin{proposition}
  The ps-contexts are linear for the relation \verb|_,_◃_|, \ie
  whenever $\Gamma$ is a ps-context, the relation \verb|Γ , _ ◃ _|
  defines a linear order on the variables of $\Gamma$.

\begin{verbatim}
ps-◃-linear : ∀ Γ → Γ ⊢ps → ◃-linear Γ
\end{verbatim}
\end{proposition}
The proof of this proposition provided in~\cite{catt} relies on
semantic consideration and the link between the category \Glob{} and
the globular sets. In our approach, we instead give a purely syntactic
proof of this result. This makes the proof very technical. The main
ingredient of the proof is a subtle invariant, which states that
whenever we have \verb|Γ ⊢ps x # A| and \verb|Γ , x ◃ y|, then
necessarily \verb|y| is an iterated target of \verb|x| in the context
\verb|Γ|.

\begin{theorem}
  The judgement \verb|_⊢ps| is decidable, and any two derivation of
  the same judgement of this form are equal.
  \[
    \begin{minipage}{.57\linewidth}
    \begin{verbatim}
is-prop-⊢ps : ∀ Γ → is-prop (Γ ⊢ps)
\end{verbatim}
    \end{minipage}
    \begin{minipage}{.43\linewidth}
\begin{verbatim}
dec-⊢ps : ∀ Γ → dec (Γ ⊢ps)
\end{verbatim}
    \end{minipage}
  \]
\end{theorem}
These results are proven by induction on the derivation trees, but
they are not straightforward. Indeed, for instance in the case of the
uniqueness, a derivation of \verb|Γ ⊢ps| necessarily comes from a
derivation of the form \verb|Γ ⊢ps x # ∗| and by induction this
derivation is necessarily unique. However the hard part is to prove
that there can only be a unique \verb|x| such that we have
\verb|Γ ⊢ps x # ∗|. Using the $\triangleleft$-linearity, we can prove
a more general lemma: if we have a derivation of \verb|Γ ⊢ps x # A|
and of \verb|Γ ⊢ps y # A| with \verb|A| and \verb|B| two types of the
same dimension, then \verb|x == y|. The decidability also presents a
difficulty: The rule \verb|psd| contains variables in its premises
that are not bound in its conclusion. However, these variables have to
belong to the context, so we can solve this issue by enumeration of
the variables and $\triangleleft$-linearity.

\subparagraph*{Index of term constructors.}
With the ps-contexts, we can define the index type for the term
constructors in the theory \CaTT{}. Recall that the term constructors
in this theory are defined by $\cohop_{\Gamma,A}$ and
$\coh_{\Gamma,A}$, where $\Gamma$ is a ps-context and $A$ is a type.
In our informal presentation, we also required the side conditions
$\Vop$ and $\Vcoh$ in the derivation rules. For convenience, we
integrate these side conditions in the index type in our
formalisation, and define \verb|J| to be the type of pairs of the form
$(\Gamma , A)$, where $\Gamma$ is a ps-context and $A$ is a type
satisfying either $\Vop$ or $\Vcoh$.

The conditions $\Vop$ and $\Vcoh$ are not straightforward to formalise
in HoTT, because the same variable may appear several times in the
same term, so there may be several witnesses that a term contains all
the desired variables. However, the intended semantics is that of a
proposition. To solve this issue, we work with the type \verb|set| of
sets of numbers, for which the membership relation is a
proposition. We define the type \verb|A ⊂ B| of witnesses that a set
\verb|A| is included in a set \verb|B|, as well as the type
\verb|A ≗ B = (A ⊂ B) × (B ⊂ A)| of set equality. We can show that
these two types are proposition since we are only manipulating finite
subsets of $\mathbb{N}$, and thus we recover the intended semantics.

\begin{definition}\label{def:isrc}
  We define the set of source variables and the set of target
  variables of a ps-context. We proceed by induction and first define
  the $i$-sources and $i$-targets by induction on the judgement
  \verb|Γ ⊢ps x # A| (we denote \verb|l| for \verb|ℓ Γ|)
\begin{verbatim}
srcᵢ-var i pss = if i ≡ 0 then nil else (nil :: 0)
srcᵢ-var i (psd Γ⊢psx) = srcᵢ-var i Γ⊢psx
srcᵢ-var i (pse {Γ = Γ} {A = A} Γ⊢psx) with dec-≤ i (S (dim A))
  ... | inl i≤SdimA = srcᵢ-var i Γ⊢psx
  ... | inr SdimA<i = (srcᵢ-var i Γ⊢psx :: l) :: (S l)

tgtᵢ-var i pss = if i ≡ 0 then nil else (nil :: 0)
tgtᵢ-var i (psd Γ⊢psx) = tgtᵢ-var i Γ⊢psx
tgtᵢ-var i (pse {Γ = Γ} {A = A} Γ⊢psx) with dec-≤ i (S (dim A))
  ... | inl i≤SdimA = if i ≡ S (dim A) then drop(tgtᵢ-var i Γ⊢psx) :: l
                                       else tgtᵢ-var i Γ⊢psx
  ... | inr SdimA<i = (tgtᵢ-var i Γ⊢psx :: l) :: (S l)
\end{verbatim}
  Here the \verb|with| construction matches on the decidability of the
  order in $\mathbb{N}$. For instances matching on \verb|dec-≤ i n|
  produces two cases of type \verb|i ≤ n| and \verb|n < i|. Moreover
  \verb|drop| takes a list and removes its head. The source and target
  sets of a ps-context are the sets corresponding to the source and
  target lists in dimensions maximal.
\begin{verbatim}
src-var (Γ , ps Γ⊢psx) = set-of-list (srcᵢ-var (dimC Γ) Γ⊢psx)
tgt-var (Γ , ps Γ⊢psx) = set-of-list (tgtᵢ-var (dimC Γ) Γ⊢psx)
\end{verbatim}
\end{definition}
\begin{definition}\label{def:src}
  We define the type \verb|_is-full-in_|, witnessing weather either
  the condition $\Vop$ or $\Vcoh$ is satisfied, as follows
\begin{verbatim}
data _is-full-in_ where
  Cop  : (src-var Γ) ≗ ((varT A) ∪-set (vart t)) →
         (tgt-var Γ) ≗ ((varT A) ∪-set (vart u)) →
         (⇒ A t u) is-full-in Γ
  Ccoh : (varC (fst Γ)) ≗ (varT A) → A is-full-in Γ
\end{verbatim}
  where \verb|varT| (\resp \verb|vart|, \verb|varC|) is the set of
  variables associated to a type (\resp to a term, to a context).
\end{definition}
\begin{definition}
  The type \verb|J = Σ (ps-ctx × Ty) λ {(Γ , A) → A is-full-in Γ}| to
  be the index type for the term constructors of the theory
  \CaTT{}. It is the types of pairs $(\Gamma, A)$ where $\Gamma$ is a
  ps-context and $A$ is a raw type satisfying $\Vop$ or $\Vcoh$.
\end{definition}
The raw syntax of the type theory \CaTT{} is the raw syntax of the
globular type theory with index type \verb|J|.

\subparagraph*{The type theory \CaTT{}}
We give a definition of the dependent type theory \CaTT, by adding the
rules to the raw syntax. For this it suffices to define a context
\verb|Ci i| a type \verb|Ti i| for every inhabitant \verb|i| of the
type \verb|J|.
\begin{definition}
  Considering a term \verb|(((Γ , Γ⊢ps), A), A-full)| of type
  \verb|J|, we pose
\begin{verbatim}
Ci (((Γ , Γ⊢ps) , A) , A-full) = (Γ , Γ⊢ps→Γ⊢ Γ⊢ps)
Ti (((Γ , Γ⊢ps) , A) , A-full) = A
\end{verbatim}
  The theory \CaTT{} is the globular type theory obtained from these
  assignations.
\end{definition}
\begin{proposition}
  The type \verb|J| has decidable equality and satisfies the technical
  condition of well-foundedness
\begin{verbatim}
eqdecJ : ∀ (x y : J) → dec (x == y)
wfJ : ∀ j → Ci j ⊢T Ti j → dimC (Ci j) ≤ dim (Ti j)
\end{verbatim}
\end{proposition}
Since this definition realises \CaTT as a particular case of a
globular type theory, it enjoys all the properties that we have
already proved for them. In particular we have already proved
\begin{theorem}
  In the theory \CaTT{} the following statements hold.
  \begin{itemize}
  \item The theory support weakening and derivability is preserved by
    the inference rules.
  \item The theory \CaTT{} defines a category with families.
  \item Every derivable judgement in \CaTT{} has a unique derivation
    tree.
  \item The sphere contexts in \CaTT{} classify the types.
  \item Type checking is decidable in the theory \CaTT{}.
  \end{itemize}
\end{theorem}

\section{Conclusion and further work}
We have presented a full formalisation of the foundational aspects of
the dependent type theory \CaTT{}. Although this dependent type theory
is quite simple, in that it does not have any definitional equality,
Proving formally all the relevant aspects that we expected turned out
to be a substantial amount of work with highly non-trivial challenges
to solve. In particular for some of the aspects such as the
decidability of type checking, the use of a proof-assistant such as
\Agda{} appears almost mandatory given the subtlety of the
arguments. Ideally, such a foundational work should be carried only
once and made accessible for future work to rely on. Unfortunately,
there is no unifying framework for every dependent type theory, that
could allow us to obtain these properties for free for all dependent
type theory, and so in practice one has to redevelop this entire
construction for every dependent type theory. The notion of globular
type theory is very limited attempt at such a framework. A more
promising approach could be to follow the work of Gylterud with the
Myott
project\footnote{\url{https://git.app.uib.no/Hakon.Gylterud/myott}}~\cite{gylterud2021defining},
and to formalise the interesting properties for a large class of
dependent type theories.

The formalisation that we have presented, and in particular the proof
for the decidability of type checking constitutes a verified
implementation of a type checker for the theory \CaTT{}. Besides, we
have developed regular implementation of such a type
checker\footnote{\url{https://github.com/thibautbenjamin/catt}} in
OCaml. However, to improve the user experience, we have defined some
meta-operations (called suspension and functorialization) on the
syntax of the theory, that we proved correct
manually~\cite{benjamin2019suspension, benjamin2020type}. However, to
avoid relying on the correctness of our implementation, the software
computes the result of these operations and checks them like any user
inputs. This leads to inefficiency in the implementation, and is not
very satisfying. A better practice would be to define and prove
formally those meta-operations, and then export the results to
executable code in order to have a natively certified implementation
of these meta-operations.

Finally, we have also defined a dependent type theory to describe
monoidal weak $\omega$-categories, that we call \MCaTT{}. This
dependent type theory is also a globular type theory, and as such it
would be valuable to formalise it as well in the framework we have
presented (there are actually two slightly different bu equivalent
formulations for this theory~\cite{benjamin2021monoidal}
and~\cite{benjamin2020type}, and only the second one corresponds to a
globular type theory as we have defined). Moreover, our understanding
of the semantics relies strongly on translations back and forth
between the theories \CaTT{} and \MCaTT{}. Such translations are
defined by induction on the syntax and are tedious to carry, defining
them formally and proving their correctness constitutes a relevant
problem to tackle in further works.

In general, giving a formulation of higher categorical results in
terms of a syntax and a dependent type theory allows to perform this
kind of reasoning that is well understood formally by
proof-assistant. We believe that it constitutes a strong asset for
higher category theory, where the complexity of the theory itself
quickly becomes a meaningful obstacle for any non-trivial exploration
by hand of the theories.



\bibliography{articles.bib}
\end{document}